\documentclass[showpacs,showkeys,prb]{revtex4}
\usepackage{amssymb}
\usepackage{amsfonts}
\usepackage{amsmath}
\usepackage{epsfig}
\usepackage{graphicx}

\providecommand{\U}[1]{\protect\rule{.1in}{.1in}}

\newcommand{\KP}{Kronig-Penney}
\newcommand{\KPperiod}{c}
\newcommand{\vzero}{V_0}
\newcommand{\vzerodimless}{u_0}
\newcommand{\ez}{{\epsilon_z}}
\newcommand{\ezdimless}{\bar{\ez}}
\newcommand{\muadim}{\bar{\mu}}
\newcommand{\Tcritic}{T_c}
\newcommand{\Eqref}[1]{Eq. (#1)}
\newcommand{\Figref}[1]{Fig. #1}

\newcommand{\GrandPot}{\Omega}

\newcommand{\ezarg}{\beta (\ez - \mu)}
\newcommand{\ezargzero}{\beta (\ez_0 - \mu)}

\newcommand{\partialmuT}{\frac{\partial \mu}{\partial T}}

\newcommand{\Tzero}{T_0}
\newcommand{\lambdazero}{\lambda_0}
\newcommand{\bosondensity}{\eta_b}
\newcommand{\gfactor}{\gamma}
\newcommand{\KPperiodzero}{\KPperiod_0}
\newcommand{\Ttilde}{\tilde{T}}

\begin{document}

\title{BEC and dimensional crossover in a boson gas within multi-slabs}
\author{O. A. Rodr\'iguez}
\affiliation{Posgrado en Ciencias F\'{\i}sicas, UNAM; Instituto de F\'{\i}sica, UNAM}
\author{M. A. Sol\'{\i}s}
\affiliation{Instituto de F\'{\i}sica, UNAM, Apdo. Postal 20-364, 01000 M\'exico D.F.,
M\'exico}


\begin{abstract}

For an ideal Bose-gas within a multi-slabs periodic structure, we report a  dimensional crossover and discuss whether a BEC transition at $T_c \neq 0$ disappears or not.    
 The multi-slabs structure is generated via a Kronig-Penney potential perpendicular to the slabs of width $a$ and separated by a distance $b$. The ability of the particles to jump between adjacent slabs is determined by the hight $V_0$    and width $b$ of the potential barrier. Contrary to what happens in the boson gas inside a zero-width multilayers case, where the critical temperature diminishes and goes up again as a function of the wall separation, here the $T_c$ decreases continuously as the potential barrier height and the cell size $a+b$ increase. We plot the surface $T_c = 10^{-6}$ showing two prominent regions in the parameters space, which  suggest a phase transition BEC-NOBEC at $T \neq 0$.  
The specific heat shows a crossover from 3D to 2D when the height of the potential or the barrier width increase, in addition to the well known peak related to the Bose-Einstein condensation.

\end{abstract}


\maketitle

\section{Introduction}

The possibility of constructing periodic 
 structures with laser light (optical crystals) 
 \cite{opticallattice} to store and study quantum gases, has 
 created the illusion that we are able to reproduce 
 in the laboratory, and in a controlled manner, 
 the properties of real systems such as helium 
 four or three in any dimension \cite{helium}; layered cuprate superconductors \cite{cuprate}; or man-made physical system such as 
 semiconductor superlattices \cite{superlattice} or tube 
 bundle superconductors \cite{tubebundle}, among 
 others. However the physical properties of a 
 constrained quantum gas come from a combination 
 of the interaction between particles plus the 
 effect of the restrictive potentials, which in our case are periodic potentials.
 In this paper we are interested in giving in 
detail the effects of the  potential on the system, 
leaving momentarily aside  the influence of the interactions, 
so both effects won't interfere with each other.

For this, we study a 3D interactionless ideal Bose gas constrained by a {\KP} potential in one direction over the whole space. Potential barriers have a magnitude $\vzero$, width $b$ and are separated 
by a distance $a$, such that the spatial period of the potential is 
$c \equiv a + b$. The barriers spread along the $z$ axis, while 
the bosons are free to move in the $x$ and $y$ directions. To obtain the particle energies we solve the 3D Schr\"{o}dinger equation by separation of variables. The 
particle energy is given by $\epsilon = \epsilon_x + \epsilon_y + \ez$, 
where $\epsilon_x = \hbar^2 k_x^2 /2m$, $\epsilon_y = \hbar^2 k_y^2 / 2m$, 
and $\ez$ is found by solving the Schr\"odinger equation for the $z$ coordinate 
with $V(z)$ the Kronig-Penney potential \cite{KP}     
\begin{equation}
    V(z) = \vzero \sum_{n=-\infty}^{\infty} \Theta[z - (n-1)(a+b) - a] \, \Theta[n(a+b) - z]
\end{equation}
where $\Theta$ is the Heaviside step function.

\begin{figure}[h]
    \begin{center}
        \includegraphics[width=3in]{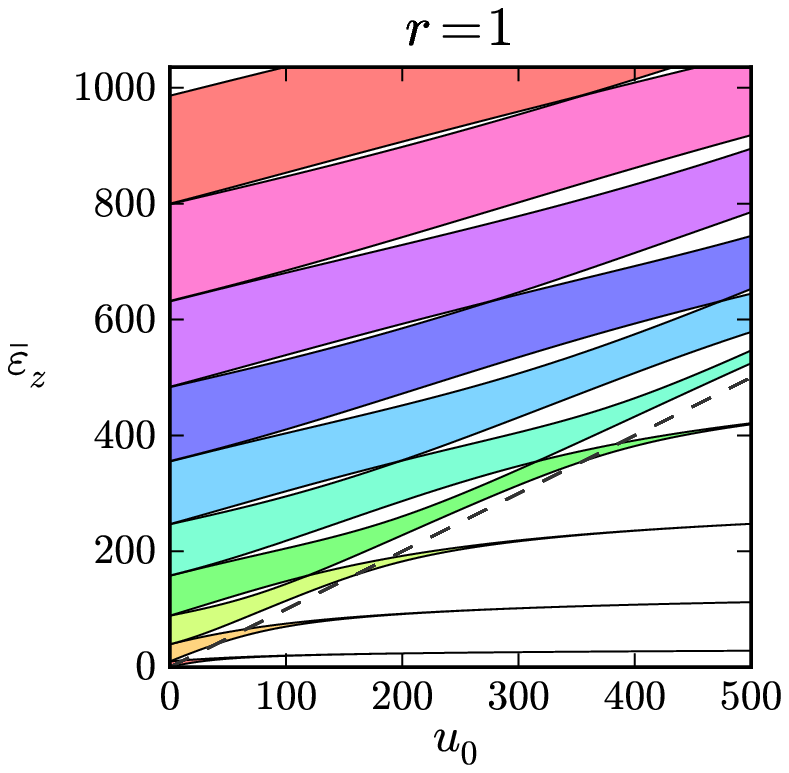}
    \end{center}
    \caption{(Color online) Energy band structure of a particle subject to a {\KP} potential. The dashed line means $\ezdimless = \vzerodimless$.}
    \label{fig:energy-spectrum-r-1}
\end{figure}

The allowed energies for a particle subject to a Kronig-Penney potential in the $z$-direction are given by equation \cite{KP}
\begin{equation}
    \label{eq:kp-dispertion-relation}
    \frac{\vzero - 2\ez}{2 \sqrt{\ez (\vzero - \ez)}} \sinh(\kappa b) \sin(\alpha a) + \cosh(\kappa b)\cos(\alpha a) = \cos(k_z (a+b))
\end{equation}
where $\kappa = \sqrt{2m(\vzero - \ez)}/\hbar$ and $\alpha = \sqrt{2m \ez}/\hbar$. 
The spectrum consist of a series of 
allowed energy bands separated by prohibited regions, where the positive side of the $j$-th band 
extends from $k_z \KPperiod = (j-1)\pi$ to $j \pi$ with $j=1,2,3,$ ... . 
 We rewrite 
\Eqref{\ref{eq:kp-dispertion-relation}} in terms
of dimensionless parameters to simplify the energy spectrum 
analysis, but not the equation itself. We use the period of the {\KP} potential $\KPperiod $ as length
unit, and define a parameter $r \equiv b/a $ which characterizes the potential by the ratio between the width of the hills and the width of the valleys in the potential. In addition, we define the dimensionless parameters
\begin{equation}
    \label{eq:dimensionless-energies}
    \vzerodimless \equiv \frac{2 m \KPperiod^2}{\hbar^2} \vzero \qquad \ezdimless \equiv \frac{2 m \KPperiod^2}{\hbar^2} \ez
\end{equation}
where the quantity $\hbar^2 / 2m\KPperiod^2$ is our unit of energy. In terms of these definitions Eq. (\ref{eq:kp-dispertion-relation}) becomes
\begin{align}
    \label{eq:kp-dispertion-relation-dimless}
    \nonumber
    \frac{\vzerodimless - 2\ezdimless}{2 \sqrt{\ezdimless (\vzerodimless - \ezdimless)}} 
    \sinh & \left( \frac{r \sqrt{\vzerodimless - \ezdimless}}{1+r} \right) 
    \sin\left( \frac{\sqrt{\ezdimless}}{1+r}\right) + \\ 
    & \cosh\left( \frac{r \sqrt{\vzerodimless - \ezdimless}}{1+r} \right)
    \cos \left(\frac{\sqrt{\ezdimless}}{1+r}\right) = \cos(k_z \KPperiod).
\end{align}
Particle energies, calculated from \Eqref{\ref{eq:kp-dispertion-relation-dimless}}, 
are displayed in \Figref{\ref{fig:energy-spectrum-r-1}} where the first ten bands of the spectrum are plotted as a function of $\vzerodimless$: regions in color
correspond to the allowed energies, the white regions are prohibited
energies. The dashed line corresponds to the energy $\ezdimless = \vzerodimless$
which serves to distinguish between energies above and below the potential magnitude.

Also from \Figref{\ref{fig:energy-spectrum-r-1}} it is clear that the energy bands below the potential height tend to 
collapse into a single energy level as $\vzerodimless$ becomes much greater than the energy of the first level of a particle in a 1D box of size $a$. A simple analysis confirms that these levels correspond to the energies of a particle in a 
box of width $a$, i.e., $ {\hbar^2 \pi^2 n^2}/{2m a^2} \ \mbox{with} \ n = \pm 1, \pm 2, \pm 3, \dots$
%
%
In other words, 
particles whose energies are much smaller than 
$\vzerodimless$ behave as particles in a box of width $a$ and infinite walls.

\section{Critical temperature}

To find the thermodynamical properties of the Bose gas we make use of
the Grand potential $\Omega$, which is given by the expression
\begin{align}
    \label{eq:grand-potential}
    \GrandPot(T, V, \mu) &= k_B T \log(1 - e^{-\ezargzero}) - \frac{m V}{(2\pi)^2 \hbar^2} \frac{1}{\beta^2} \int_{-\infty}^{\infty} dk_z \, g_2(e^{-\ezarg})
\end{align}
with $g_\sigma(z) \equiv \sum_{l=1}^{\infty} z^l / l^\sigma$ the Bose  
function \cite{Path}.
Then the number of bosons, the internal energy and the
isochoric specific heat are given in terms of the Grand potential by
\begin{equation}
    \label{eq:thermo-props}
    N = -\left(\frac{\partial \GrandPot}{\partial \mu}\right)_{T, V}; \ \ 
    U = -k_B T^2 \left( \frac{\partial}{\partial T} \left(\frac{\GrandPot}{k_B T}\right) \right)_{V, z = e^{\beta \mu}} ; \ \ 
     C_V = \left(\frac{\partial U}{\partial T} \right)_{N, V} 
\end{equation}
respectively. In order to find the Bose-Einstein critical temperature $\Tcritic$ we begin with the number equation
\begin{equation}
    \label{eq:number-equation}
    N = \frac{1}{e^{\ezargzero} -1} - \frac{m V}{(2\pi)^2 \hbar^2} \frac{1}{\beta}  \int_{-\infty}^{\infty} dk_z \, \log(1 - e^{-\beta (\ez - \mu)}) 
\end{equation}
where the integrals over $x$ and $y$ directions have been done. The separation between the bosons in the ground state $\ez_0$, namely the condensed bosons
$N_0(T)$, 
and those in the excited states $N_{e}(T)$, i.e. 
$N = N_0 + 
N_{e}$, becomes evident. The Bose-Einstein condensation occurs at $T = \Tcritic$ when all the particles are in the excited states, i.e., $N \approx N_{e}$, and the bosons are pushed \emph{en masse} to the ground state. At this temperature the chemical
potential corresponds to the ground energy of the system, $\mu(\Tcritic) = \mu_0 = \ez_0$, so
\begin{equation}
    \label{eq:critical-temperature-equation}
    N = - \frac{m V}{(2\pi)^2 \hbar^2} \frac{1}{\beta_c}  \int_{-\infty}^{\infty} dk_z \, \log(1 - e^{-\beta_c (\ez - \mu_0)})
\end{equation}
with $\beta_c = 1/k_B \Tcritic$. This expression determines $\Tcritic$ in an
implicit way. Our system has infinite size and an infinite number of 
bosons but the particle density $\bosondensity = N/V$ is a constant.

\begin{figure}[h]
    \begin{center}
        \includegraphics[width=3in]{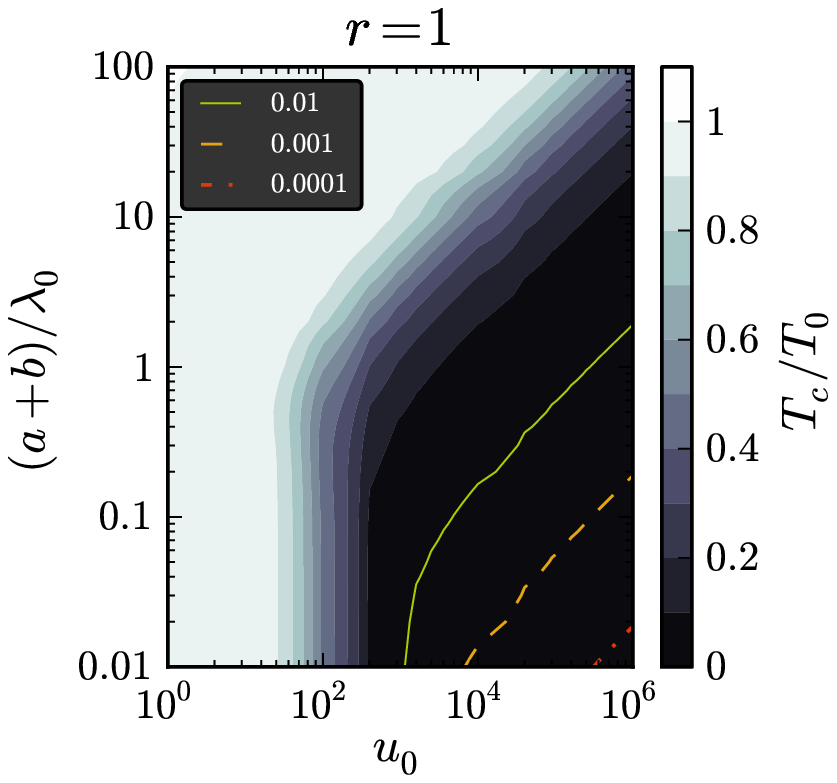}
    \end{center}
    \caption{(Color online) Contour levels of \ $\Tcritic/\Tzero$}
    \label{fig:tc-contours-2D}
\end{figure}

\begin{figure}[h]
    \begin{center}
        \includegraphics[width=3in]{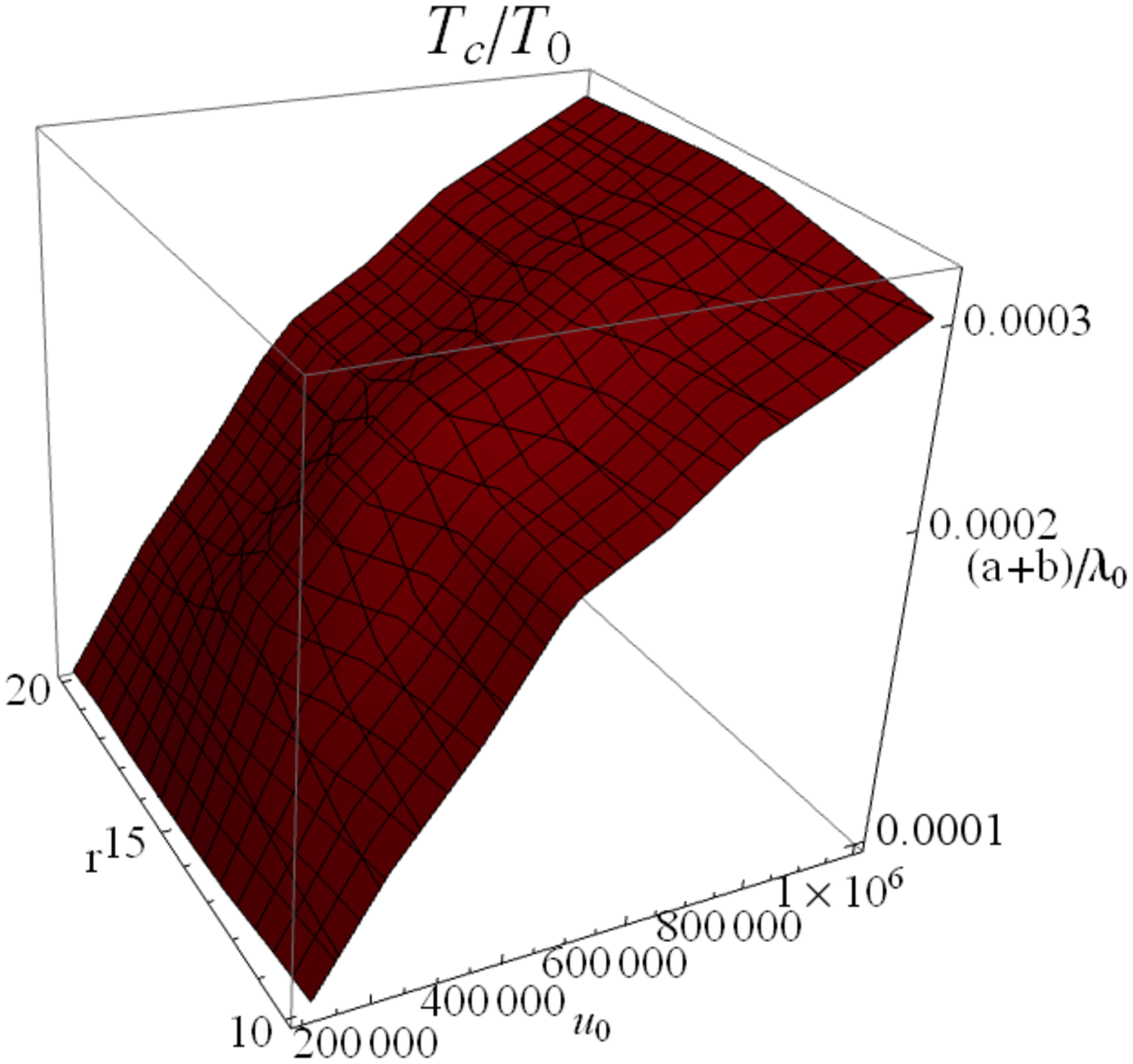}
    \end{center}
    \caption{(Color online) Surface of $\Tcritic/\Tzero = 10^{-6}$}
        \label{fig:tc-contour-3D}
\end{figure}

As we did with the energy equation (\ref{eq:kp-dispertion-relation}) it is better to work with dimensionless parameters in order to get numerical results from Eq. (\ref{eq:critical-temperature-equation}). We define a parameter 
$\gfactor \equiv \hbar^2/2m\KPperiod^2 k_B \Tzero$ as the quotient of two energies,
one of them being the unit of energy used in \Eqref{\ref{eq:dimensionless-energies}}, the other being a thermal energy, 
$k_B \Tzero$, where $\Tzero$ is the critical temperature of an infinite free ideal boson gas with density $\bosondensity$
\begin{equation}
    \Tzero = \frac{2 \pi \hbar^2}{m k_B} \left( \frac{\bosondensity}{\zeta(3/2)} \right)^{2/3} \approx 3.31 \frac{\hbar^2}{m k_B} \bosondensity^{2/3}
\end{equation}
and $\zeta(u)$ is the Riemann Zeta function.  It turns out that $\gfactor$ can
be rewritten in terms of the parameter $\KPperiodzero = \KPperiod/\lambdazero$
as $\gfactor = 1/(4 \pi \KPperiodzero^2)$, so $\KPperiodzero$ is the 
dimensionless parameter that relates the period of the potential with the
thermal wavelength $\lambdazero$ of an ideal boson gas at its Bose-Einstein condensation temperature $\Tzero$. So with the definitions of $\gfactor$, 
$\Tzero$ and $\KPperiodzero$ the number equation becomes
\begin{equation}
-1 = \frac{2 \sqrt{\gfactor/\pi}}{\zeta(3/2)} \Ttilde_c \int_{0}^{\infty} dk'_z \, \log(1 - e^{-\gfactor (\ezdimless - \muadim_0) / \Ttilde_c})
\end{equation}
with $\Ttilde_c = T/\Tcritic$ and $dk'_z = \KPperiod dk_z$. In this expression
the energies used are those obtained from \Eqref{\ref{eq:kp-dispertion-relation-dimless}}, where the parity of the energy 
over $k_z$ has been used. 
The critical temperature $\Tcritic$ in units of $\Tzero$, for $r=1$, are shown
in \Figref{\ref{fig:tc-contours-2D}} as 2D contours plot, where darker colors 
represent lower critical temperatures. \Figref{\ref{fig:tc-contour-3D}} shows 
the isosurface $\Tcritic/\Tzero = 10^{-6}$ for a 3D interval over 
$\vzerodimless$, $r$ and $(a+b)/\lambdazero$.

We can see that the direct result of increasing $\vzerodimless$ is to reduce
$\Tcritic$, as the barriers tend to trap the bosons in narrow bands for 
$\ezdimless$ while they are free to move in the other two directions. This
means that the systems behave more like a 2D boson gas as $\vzerodimless$
grows, with the corresponding reduction of $\Tcritic$ towards zero. At the
same time as the period $\KPperiod$ increases 
the critical temperature $\Tcritic$ rises
as the valley regions between the barriers grow, reducing the effect of the barrier potential on the 
bosons and returning the system to a 3D free gas with 
$\Tcritic = \Tzero$. A reduction of the period does not increase $\Tcritic$,
it keeps it constant in contrast to the zero-width multilayer potential \cite{Paty} case where a period reduction, below a critical value,  increases the critical temperature .

\section{Isochoric specific heat}

After some algebra we obtain an expression for the {\it isochoric specific heat} $C_V$ from the internal
energy of the system as stated in \Eqref{\ref{eq:thermo-props}}
\begin{align}
    \nonumber
    C_V &= 
    -\frac{m V}{(2\pi)^2 \hbar^2} \frac{1}{\beta T} \int_{-\infty}^{\infty} dk_z \, (\ez - \ez_0) \log(1 - e^{-\ezarg}) \\
    \nonumber
    &+ \frac{m V}{(2\pi)^2 \hbar^2} \frac{2}{\beta^2 T} \int_{-\infty}^{\infty} dk_z \, g_2(e^{-\ezarg}) \\
    \nonumber
    &+ \frac{m V}{(2\pi)^2 \hbar^2} \frac{1}{T} \int_{-\infty}^{\infty} dk_z \, (\ez - \ez_0) \frac{(\ez - \mu) + T \partialmuT}{e^{\ezarg} - 1} \\
    \label{eq:specific-heat-equation}
    & - \frac{m V}{(2\pi)^2 \hbar^2} \frac{1}{\beta T} \int_{-\infty}^{\infty} dk_z \left( (\ez - \mu) + T \partialmuT \right) \log(1 - e^{-\ezarg})
\end{align} 
which depends on the chemical potential $\mu$ and its derivative $\left( \partial \mu / \partial T \right)_{N, V}$. For $T \leq \Tcritic$ $\mu = \mu_0$ and $\left( \partial \mu / \partial T \right)_{N, V} = 0$.
\begin{figure}
    \begin{center}
        \includegraphics[width=4in]{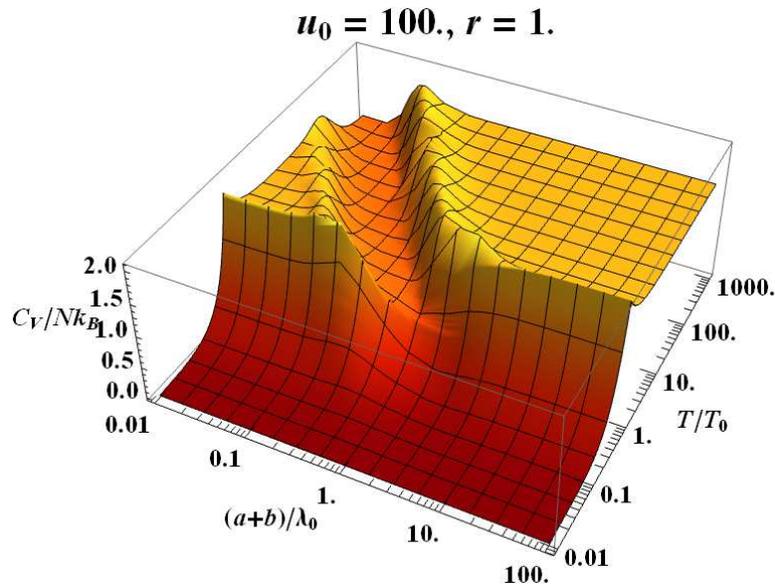}    
    \end{center}
    \caption{(Color online) Specific heat per boson}
    \label{fig:cv-surface}
\end{figure}

In \Figref{\ref{fig:cv-surface}} we show a 3D surface of the specific heat per boson $C_V/N k_B$ as a function
of $(a+b)/\lambdazero$ and $T/\Tzero$, for $u_0 = 100$ and $b/a = 1$.   It show 
the classic two-phase behavior of $C_V$ for a boson gas, increasing monotonically up to 
$T = \Tcritic$ where it develops a peak,
but surprisingly not 
decreasing monotonically to the classical value of $3/2$ as $T$ grows. For temperatures larger than the critical temperature, i.e., in the normal phase, 
there is a complex behavior
with one local minimum and one or two
local maximums depending on $\KPperiod/\lambdazero$ value. In particular, we have 
found that the local minimum is directly related to the thermal wavelength of
the bosons $\lambda = h/\sqrt{2m \pi k_B T}$, being approximately twice the 
size of the spatial period of the potential. More specifically, the bosons are
effectively trapped by the potential in the region where $C_V$ is minimal, 
behaving as a 2D gas. The value of $C_V$ in this regions drops to one, as
it is expected for a 2D classic gas. 
On the other hand,
the extension of the range $T/\Tzero$ where $C_V \approx 1$ 
depends on $\vzerodimless$ and $\KPperiod/\lambdazero$ values. For large values of $u_0$ the minimum becomes a plateau and for some 
combinations of these parameters the drop does not even exist.
By instance, this complex behavior is not present for large $\KPperiod/\lambdazero$ values because under this condition the 
interaction of the potential with the bosons 
reduces and the gas seems more like a 3D gas. However, starting from some critical value of $\KPperiod/\lambdazero$
the minimum is present for 
every value as $\KPperiod/\lambdazero \to 0$, and moves to higher temperatures.
It is worth noting that $T$ is displayed
in a logarithmic scale, so the temperatures required to brake the 2D behavior could be very large in relation to $\Tzero$.

\section{Conclusions}

In summary, we have calculated the Bose-Einstein critical temperature and the isochoric specific heat, of a Bose gas within an infinite stack of linked slabs. We found that the 
critical temperature decreases monotonically as the potential magnitude increase, keeping constant the cell size of the KP potential. In addition, the critical temperature tends to  
a fixed value smaller than $T_0$ as the period decreases which contrast with that behavior observed for bosons within layers of zero width. Both facts show that we can always find a region where the critical temperature is as small as we wish given the adequate parameters but without a clear signature of a transition to $T_c =0$.
On the other hand, the specific heat shows a complex behavior. For some regions of $(a+b)/\lambda_0$ is clear the existence of a Bose-Einstein condensation which is manifested by a peak. In the normal phase $T > T_c$, appears two local maxima and one minimum. The minimum is related to the complete boson trapping by the barriers as the boson wavelength is  approximately equal to twice the spatial unit cell. When the $u_0$ value overcome a critical value, the minimum becomes a plateau around the classical value of a 2D system.
This behavior disappears at 
higher temperatures where the particle are able to overcome the height of potential or when the potential period   becomes much larger than
 $\lambdazero$, as expected, and the specific heat value goes to the classical value of 3/2. 

{\bf Acknowledgements} We acknowledge partial support from DGAPA UNAM through the projects IN-105011 and IN-111613.


\end{document}